\documentclass[a4paper]{spie}  

\usepackage{amsmath,amsfonts,amssymb}
\usepackage{graphicx}
\usepackage[colorlinks=true, allcolors=blue]{hyperref}
\usepackage{verbatim}
\title{
Self-Supervised U-Net for Segmenting Flat and Sessile Polyps \\[1ex] \large Accepted in SPIE Medical Imaging Symposium 2021 
}

\author[a]{Debayan Bhattacharya}
\author[b]{Christian Betz}
\author[c]{Dennis Eggert}
\author[d]{Alexander Schlaefer}
\affil[a,d]{Hamburg University of Technology, Hamburg, Germany}
\affil[a,b,c]{ Universitätsklinikum Hamburg-Eppendorf, Hamburg, Germany}

\authorinfo{Send correspondence to debayan.bhattacharya@tuhh.de if required}

\begin{document} 
\maketitle

\section{PURPOSE}
\label{sec:intro}  

Colorectal Cancer(CRC) poses a great risk to public health. It is the third most common cause of cancer in the US \cite{pmid31912902}. Development of colorectal polyps is one of the earliest signs of cancer. Early detection and resection of polyps can greatly increase survival rate to 90\%\cite{2016benchmark}. Manual inspection can cause misdetections because polyps vary in color, shape, size and appearance. To this end, Computer-Aided Diagnosis systems(CADx) has been proposed that detect polyps by processing the colonoscopic videos\cite{Suzuki2012}. The system acts a secondary check to help clinicians reduce misdetections so that polyps may be resected before they transform to cancer. 

Polyps vary in color, shape, size, texture and appearance. As a result, the miss rate of polyps is between 6\% and 27\%\cite{82d6bf73e5204a82933c7d7d69c04735} despite the prominence of CADx solutions. Furthermore, sessile and flat polyps which have diameter less than 10 mm are more likely to be undetected \cite{Zimmermann-Fraedrich2019}. Convolutional Neural Networks(CNN) have shown promising results in polyp segmentation\cite{10.1007/978-3-319-24574-4_28,patel2021enhanced,jha2019resunet}. However, all of these works have a supervised approach and are limited by the size of the dataset. It was observed that smaller datasets reduce the segmentation accuracy of ResUNet++\cite{9314114}. 

Self-supervision is a stronger alternative to fully supervised learning especially in medical image analysis since it redresses the limitations posed by small annotated datasets. From the self-supervised approach proposed by Jamaludin \textit{et al.}\cite{jamaludin2017selfsupervised}, it is evident that pre-training a network with a proxy task helps in extracting meaningful representations from the underlying data which can then be used to improve the performance of the final downstream supervised task.

In summary, we train a U-Net to inpaint randomly dropped out pixels in the image as a proxy task. The dataset we use for pre-training is Kvasir-SEG\cite{jha2020kvasir} dataset. This is followed by a supervised training on the limited Kvasir-Sessile\cite{jha2020sessile} dataset. Our experimental results demonstrate that with limited annotated dataset and a larger unlabeled dataset, self-supervised approach is a better alternative than fully supervised approach. Specifically, our self-supervised U-Net performs better than five segmentation models which were trained in supervised manner on the Kvasir-Sessile dataset.

\section{METHODOLOGY}

\subsection{Datasets}

\begin{itemize}
    \item Kvasir-SEG \cite{jha2020kvasir} contains 1000 polyp images and their segmentatation masks. The size of the images vary between 332x487 and 1920x1072 pixels. Every image contains atleast one polyp. 
    \item Kvasir-Sessile \cite{jha2020sessile} contains 196 polyps with size smaller than 10 mm. Expert gastroenterologists have selected the flat and sessile polyps. This dataset is a subset of the Kvasir-SEG dataset. 

 For our supervised task, we randomly pick  10\% of the Kvasir-Sessile dataset to make the test set. From the remaining images, we make five folds of equal sizes and leave-one-fold-out for cross validation. The number of training, testing and validation images used for evaluating our models are shown in Table \ref{tab:dataset}. We remove all the images from our pre-training dataset that are in the test set of our downstream supervised task. As a result, we use 980 images of Kvasir-SEG for pre-training.     

\end{itemize}

\begin{table}[!ht]
\centering
\caption{ Datasets used in our experiments}

\begin{tabular}{|c| c | c |c |c |c|} 

 \hline
 Dataset & Used for &Train images & Validation Images & Test Images & Input Size\\ [0.5ex] 
 \hline
 Kvasir-SEG & Pre-training    & 980  & - & - & Variable \\ 
 Kvasir-Sessile & Supervised Training  & 140 & 36 & 20 & Variable \\  [1ex] 
 \hline
 
\end{tabular}

\label{tab:dataset}

\end{table}
\subsection{Proxy Inpainting Task}

 Inpainting of images is a technique where the source image has pixels which are randomly dropped out from it. The encoder of the model encodes the context of the image and compresses into a latent representation. The decoder uses this latent representation to fill in the missing pixels observed in the image. In doing so, the features learned by the model consider the color, shape, size and various other attributes of the input image distribution.  We pre-train our U-Net with an L2 reconstruction loss shown below: 

\begin{equation}
    L_{rec}(x) = \left \| M \odot (x - F((1-M) \odot x,\theta)) \right \|_{2}^{2}
\end{equation}
Here, M is a binary mask where the dropped image pixel has value 1 and it has value 0 wherever the pixel is not dropped. The model is represented by \(F(.)\) and parameterised by \(\theta\). \(\odot\) represents element-wise product operation and \(x \epsilon \mathbb{R}^{H \times W \times 3} \) is the input image.

\begin{figure}
    \centering
    \includegraphics[width=1\textwidth]{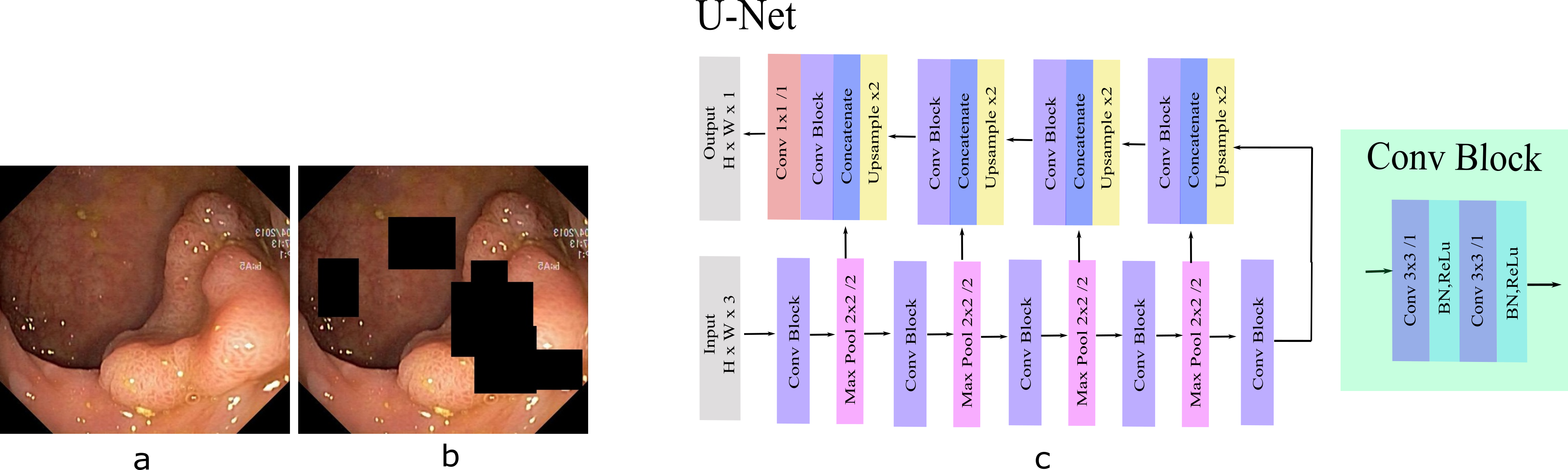}
    \caption{(a) Original Image (b) Image with random pixel dropout (c) U-Net architecture. BN denotes Batch Normalisation. 2x2 and 3x3 denote kernel sizes. /1 and /2 denote stride taken by the kernels. Upsample operation is binary interpolation. x2 denotes the scale of upsampling.}
    \label{fig:proxy}
\end{figure}

\subsection{Models for segmentation of flat and sessile polyps}

We train five segmentation models in a fully supervised manner. The models used in this work are U-Net\cite{10.1007/978-3-319-24574-4_28}, Attention U-Net\cite{oktay2018attention}, R2U-Net\cite{alom2018recurrent}, R2AU-Net\cite{Zuo2021} and ResUNet++\cite{jha2019resunet}. We have used Tversky loss\cite{salehi2017tversky} for all the supervised training experiments with  \(\alpha\) = 0.4 and \(\beta\) = 0.6 found through grid search. 

\section{EXPERIMENTS AND RESULTS}
\subsection{Implementation Details}

 \subsubsection{Pre-training and Supervised Experiments}
 We train all the models for 65 epochs with a batch size of 4. Only the U-Net is pre-trained with the Kvasir-SEG dataset. Supervised training is accomplished with Kvasir-Sessile dataset. For pre-training, the first 50 epochs are trained with a learning rate of 1e-5 and the remaining 15 epochs are trained with 1e-6. For supervised training, the first 50 epochs are trained with a learning rate of 1e-4 and the remaining 15 epochs are trained with 1e-5. We use data augmentation such as random flipping and random rotation. The pre-training step has an additional random pixel dropout augmentation as shown in Figure \ref{fig:proxy} (b). Each batch of images are trained with sizes 192x192, 320x320 and 512x512. Adam optimiser with default parameters has been used for training. We use PyTorch for all our experiments.

 \subsubsection{Evaluation}
 To prove the generalisation and performance improvement from pre-training, we do a five-fold cross validation experiment for all the models and use dice coefficient(DSC), mean IoU(mIoU), Precision and Recall to evaluate the models.

\subsection{Results}
 
\begin{figure}
    \centering
    \includegraphics[width=1\textwidth]{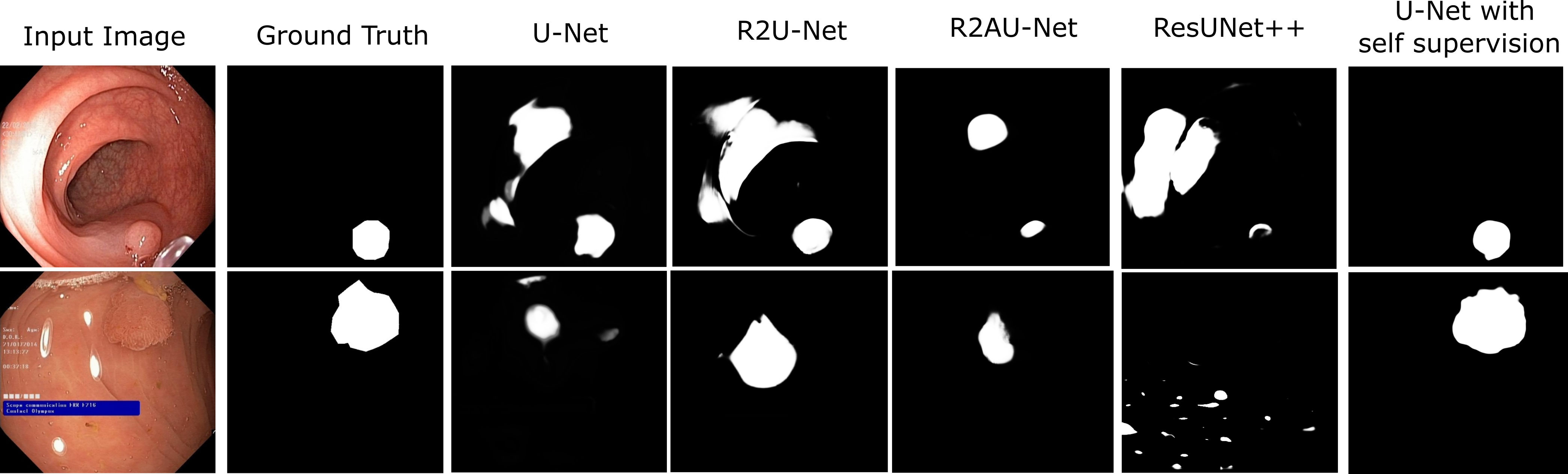}
    \caption{Qualitative analysis of predicted segmentation maps}
    \label{fig:qualitative}
\end{figure}

From Table \ref{tab:results}, we observe that self-supervised model consistently outperforms supervised models in all the metrics. In particular, we see U-Net with self-supervision increases the DSC by 0.29, 0.31, 0.32, 0.36 and 0.30 in comparison to DSC of fully supervised U-Net, Attention U-Net, R2U-Net, R2AU-Net and ResUNet++ respectively. Similarly, the mIoU has an improvement of 0.29, 0.31, 0.31, 0.34 and 0.29 increase relative to  U-Net, Attention U-Net, R2U-Net and ResUNet++ respectively. The precision improves by 0.31, 0.39, 0.35, 0.36 and 0.35 relative to  U-Net, Attention U-Net, R2U-Net, R2AU-Net and ResUNet++ respectively. Finally, the recall shows an improvement of 0.21, 0.18, 0.28, 0.35 and 0.2 relative to U-Net, Attention U-Net, R2U-Net, R2A-UNet and ResUNet++ respectively. Another observation to be made is the smaller DSC and mIoU of the larger models in comparison to the U-Net. The qualitative analysis of our findings are presented in Figure \ref{fig:qualitative}.
 
\begin{table}[!ht]
\centering
\caption{Experimental results}
\begin{tabular}{|c| c |c |c |c|} 

\hline
 Method & DSC & mIoU & Precision & Recall\\ [0.5ex] 
 \hline
 U-Net     & $0.31\pm0.03$   & $0.20\pm0.03$ & $0.42\pm0.04$ & $0.48\pm0.04$ \\ 
 Attention U-Net  & $0.29\pm0.04$ & $0.18\pm0.03$ & $0.34\pm0.04$ & $0.51\pm0.05$ \\  
 R2U-Net  & $0.28\pm0.05$ & $0.18\pm0.04$ & $0.38\pm0.08$ & $0.41\pm0.06$ \\  
 R2AU-Net & $0.24\pm0.05$ & $0.15\pm0.03$ & $0.37\pm0.03$ & $0.34\pm0.08$ \\ 
 ResUNet++  & $0.30\pm0.03$ & $0.20\pm0.03$ & $0.38\pm0.05$ & $0.49\pm0.04$ \\

  U-Net (with self-supervision) & $\textbf{0.60}\pm\textbf{0.05}$ & $\textbf{0.49}\pm\textbf{0.04}$ & $\textbf{0.73}\pm\textbf{0.06}$ & $\textbf{0.69}\pm\textbf{0.01}$\\ [1 ex]
 \hline
 
\end{tabular}

\label{tab:results}

\end{table}

\section{NEW OR BREAKTHROUGH WORK TO BE PRESENTED}

To the best of our knowledge, this is the first use of self-supervision in polyp segmentation task. Prior works\cite{patel2021enhanced,jha2019resunet} focused on fully supervised approaches wherein they modified the baseline U-Net and incorporated feature enhancement modules to strengthen the feature representation of the network. We demonstrate that pre-training a basic U-Net that is devoid of feature enhancement modules achieves superior performance in comparison to five existing segmentation models. We observe that challenging polyps can be detected more precisely with very limited annotated data if pre-training is done. Furthermore, we see that models with more trainable parameters perform poorer than U-Net with full supervision. Large models tend to overfit on small annotated datasets. Our results reflect this scenario. This also highlights the disadvantage of using large models with small annotated datasets. In conclusion, we believe that self-supervision is an important learning paradigm to make lightweight models generalise better to unseen data. This is particularly important if we want to implement Artificial Intelligence driven CADx in clinical practice where  computing resources are not always present. We hope our work will aid future research in this direction. 

\section{CONCLUSION} 
We have shown that self-supervised U-Net in polyp segmentation task performs better than fully supervised models for the same training dataset. We compare our model against popular segmentation models that have feature enhancement modules(exception: U-Net). In particular, we show that segmenting challenging polyps that are flat can be segmented more precisely if we do self-supervision. We perform five-fold cross validation experiments with all models to show the generalisation and robustness of our approach. Our methodology is general which leaves area for research in other medical imaging domains. 

\section{ACKNOWLEDGMENTS}
The authors have no conflicts of interests to report. This work has not been submitted for publication anywhere else.
\bibliography{report} 

\begin{thebibliography}{10}

\bibitem{pmid31912902}
Siegel, R.~L., Miller, K.~D., and Jemal, A., ``{{C}ancer statistics, 2020},''
  {\em CA Cancer J Clin}~{\bf 70},  7--30 (01 2020).

\bibitem{2016benchmark}
Vázquez, D., Bernal, J., Sánchez, F.~J., Fernández-Esparrach, G., López,
  A.~M., Romero, A., Drozdzal, M., and Courville, A., ``A benchmark for
  endoluminal scene segmentation of colonoscopy images,'' (2016).

\bibitem{Suzuki2012}
Suzuki, K., ``A review of computer-aided diagnosis in thoracic and colonic
  imaging,'' {\em Quantitative imaging in medicine and surgery}~{\bf 2},
  163--176 (Sep 2012).
\newblock 23256078[pmid].

\bibitem{82d6bf73e5204a82933c7d7d69c04735}
Ahn, S., Han, D., Bae, J., Byun, T., Kim, J., and Eun, C., ``The miss rate for
  colorectal adenoma determined by quality-adjusted, back-to-back
  colonoscopies,'' {\em Gut and liver}~{\bf 6},  64--70 (Jan. 2012).

\bibitem{Zimmermann-Fraedrich2019}
Zimmermann-Fraedrich, K., Sehner, S., Rex, D.~K., Kaltenbach, T., Soetikno, R.,
  Wallace, M., Leung, W.~K., Guo, C., Gralnek, I.~M., Brand, E.~C., Groth, S.,
  Schachschal, G., Ikematsu, H., Siersema, P.~D., and R{\"o}sch, T.,
  ``Right-sided location not associated with missed colorectal adenomas in an
  individual-level reanalysis of tandem colonoscopy studies,'' {\em
  Gastroenterology}~{\bf 157},  660--671.e2 (Sep 2019).

\bibitem{10.1007/978-3-319-24574-4_28}
Ronneberger, O., Fischer, P., and Brox, T., ``U-net: Convolutional networks for
  biomedical image segmentation,''  234--241, Springer International
  Publishing, Cham (2015).

\bibitem{patel2021enhanced}
Patel, K., Bur, A.~M., and Wang, G., ``Enhanced u-net: A feature enhancement
  network for polyp segmentation,'' (2021).

\bibitem{jha2019resunet}
Jha, D., Smedsrud, P.~H., Riegler, M.~A., Johansen, D., de~Lange, T.,
  Halvorsen, P., and Johansen, H.~D., ``Resunet++: An advanced architecture for
  medical image segmentation,'' (2019).

\bibitem{9314114}
Jha, D., Smedsrud, P.~H., Johansen, D., de~Lange, T., Johansen, H.~D.,
  Halvorsen, P., and Riegler, M.~A., ``A comprehensive study on colorectal
  polyp segmentation with resunet++, conditional random field and test-time
  augmentation,'' {\em IEEE Journal of Biomedical and Health Informatics}~{\bf
  25}(6),  2029--2040 (2021).

\bibitem{jamaludin2017selfsupervised}
Jamaludin, A., Kadir, T., and Zisserman, A., ``Self-supervised learning for
  spinal mris,'' (2017).

\bibitem{jha2020kvasir}
Jha, D., Smedsrud, P.~H., Riegler, M.~A., Halvorsen, P., de~Lange, T.,
  Johansen, D., and Johansen, H.~D., ``Kvasir-seg: A segmented polyp dataset,''
  in [{\em International Conference on Multimedia
  Modeling}{\nolinebreak\hspace{0.1em}]},   451--462, Springer (2020).

\bibitem{jha2020sessile}
Jha, D., Smedsrud, P.~H., Johansen, D., de~Lange, T., Johansen, H.~D.,
  Halvorsen, P., and Riegler, M.~A., ``A comprehensive study on colorectal
  polyp segmentation with resunet++, conditional random field and test-time
  augmentation,'' (2020).

\bibitem{oktay2018attention}
Oktay, O., Schlemper, J., Folgoc, L.~L., Lee, M., Heinrich, M., Misawa, K.,
  Mori, K., McDonagh, S., Hammerla, N.~Y., Kainz, B., Glocker, B., and
  Rueckert, D., ``Attention u-net: Learning where to look for the pancreas,''
  (2018).

\bibitem{alom2018recurrent}
Alom, M.~Z., Hasan, M., Yakopcic, C., Taha, T.~M., and Asari, V.~K.,
  ``Recurrent residual convolutional neural network based on u-net (r2u-net)
  for medical image segmentation,'' (2018).

\bibitem{Zuo2021}
Zuo, Q., Chen, S., and Wang, Z., ``R2au-net: Attention recurrent residual
  convolutional neural network for multimodal medical image segmentation,''
  {\em Security and Communication Networks}~{\bf 2021},  6625688 (Jun 2021).

\bibitem{salehi2017tversky}
Salehi, S. S.~M., Erdogmus, D., and Gholipour, A., ``Tversky loss function for
  image segmentation using 3d fully convolutional deep networks,'' (2017).

\end{thebibliography}
\bibliographystyle{spiebib} 

\end{document}